\documentclass[aps,prb,reprint,groupedaddress]{revtex4-1}
\usepackage[dvips]{graphicx}

\begin{document}
\title{Numerical Study on Spontaneous Symmetry Breaking in a XY Quantum Antiferromagnet on a Finite
Triangular Lattice }

\author{Tomo Munehisa and Yasuko Munehisa}
\email{munehisa@yamanashi.ac.jp}
\affiliation{Faculty of Engineering, University of Yamanashi}

\date{\today}



\begin{abstract}
Motivated by recent experiments that require more complicated macroscopic wave functions
in   the condensed matters,
we make 
 numerical study on  a XY quantum antiferromagnet
on a finite triangular lattice using  the variational Monte Carlo method and the  stochastic state
selection method.
One of our purpose is  a numerical confirmation on dominance of a Nambu-Goldstone boson in low energy
excitation.  For  another purpose,  
we calculate energy,   an expectation value of a symmetry breaking operator
and structure functions of spin by fixing
 a  quantum number of the symmetry. These calculations are  made 
for states that become  degenerate in an infinitely large lattice.

By numerical calculations  we  confirm existence of a Nambu-Goldstone boson, and
find dependence of a square of the quantum number for the above quantities.
 Using these results we can discuss on complicated macroscopic  wave functions in quantum spin systems.
\end{abstract}
\maketitle
\section{Introduction}\label{sec1}

It is well known that the ground states of many quantum antiferromagnets  on two dimensional lattices exhibit  
semi-classical Neel order\cite{rev},
which  has been supported strongly by the spin wave theory(SWT)\cite{sw} as well as numerical works\cite{book1,book2}.
This order implies that the ground state  is a coherent state that consists of
highly degenerated states.  
Also this ground state is characterized by  order parameters, which correspond  with a magnitude and
a phase of a macroscopic wave function.

However, experimental works in other condensed matters give us  more complicated phenomena.
For examples  experiments with alkali atoms have realized that two or more Bose-Einstein condensates 
with different phases merged
and produced an interference pattern in their densities\cite{alkali1}.
This interference pattern  has forced us to examine again theoretical descriptions
based on the coherent state\cite{BEC1,BEC2,BEC3}.
A work on superconductors \cite{MQC1} can be refereed as another experiment on 
 the interference between macroscopic wave functions.
By these experiments we have to  recognize that the ground state or the macroscopic wave function in
the condensed matters,  where
spontaneous symmetry breaking(SSB) occurs,
can be not described only by a few parameters. 

Although above experiments have not been done yet for quantum spin systems, at least many-body systems,
 theoretical investigations  are needed for the ground states in these systems if we apply more complicated  external interactions.
For these investigations  we focus our study on an effect due to finiteness of a system and a Nambu-Goldstone(NG) boson.
In a finite lattice we does not have degenerate states, but have a state with the lowest energy for
a fixed quantum number of a continuous symmetry. 
We would like to make numerical study for this state with the fixed quantum number.
By this study we can make a theoretical discussion on the ground state
with the complicated interaction.
Also cluster experiments of quite small sizes\cite{cluster1} are  another motivation of this study on a finite
system.

Although we have  calculated energy of the NG boson by the SWT \cite{sw},
we have to assume that the SSB occurs for an application of the SWT.
While in numerical works such as a Monte Carlo method,  we do not need this assumption.
But a confirmation of a NG boson  in these works is not an  easy task because
 a NG boson exists  only for a small wave vector whose calculation requires
a quite large lattice. 
Our numerical study will  be made on  a 108 site lattice, where the smallest magnitude of 
wave vectors is $2\pi/9$, 
so that we can present quantitative discussions on the SSB of quantum spin systems.

We calculate  energy,  an expectation value of a symmetry breaking operator
and structure functions of spin, which give us knowledge on the NG boson.
In order to make clear our calculations, we  introduce  notations.
If a Hamiltonian $\hat{H} $ in the system has a continuous symmetry, we define a charge operator $\hat{Q}$
whose quantum number can be fixed for the eigen state of the $\hat{H}$.
We define a state $\mid n \rangle $ for a system with a site number $N$.
$$  \hat{H} \mid n \rangle =\mid n \rangle E(N,n)  \  ,$$
$$  \hat{Q} \mid n \rangle =\mid n \rangle n \  .$$
Here $  E(N,n) $ is the lowest eigenvalue for a fixed charge number $n$.
These $ \mid n \rangle$'s become  degenerate when a lattice size is infinitely large.
Also in a finite lattice for many antiferromagnet systems, one could not see degenerated states for
the lowest energy so that we have only one state of  the lowest energy.
%
When we add an external operator $\hat{B}$ to the $\hat{H}$,
by  a modified Hamiltonian $\hat{H}_B=\hat{H} + \hat{B} $,
we obtain  an eigen state $\mid E(N,B) \rangle $ of the lowest energy $ E(N,B) $ of $\hat{H}_B$.
Here  $\mid E(B) \rangle = \sum_n \mid n\rangle c_n(B)  $.
If we control  a form of $\hat{B}$ and its magnitude,
 we have  the state of $ \mid E(N,B) \rangle $ with various coefficients $ c_n(B) $.
$ c_n(B) $ is determined by $ E(N,n) $ and $ \langle n \mid  \hat{B}  \mid n' \rangle $.

In usual experiments it is difficult to control $\hat{B}$ so that
we assume that $\hat{B}=h \sum_i (\hat{\sigma}_+(i)+\hat{\sigma}_-(i) )$,
 where $\langle n+1 \mid \hat{\sigma}_+(i)  \mid n \rangle = v \not= 0$ and
$\hat{\sigma}_-(i)=\hat{\sigma}_+(i)^\dagger $,  so that we have the coherent 
ground state where the coefficient $ c_n(B) $ is $C exp( -i n \theta ) $.
However, recent  experiments\cite{alkali1}
have told us that one can control $\hat{B}$ .
Especially in a superconducting single-electron transistor \cite{MQC1}
the ground state has only a few $\mid n \rangle $'s.
These experiments ask us a following question;
What kind of $\hat{B}$ induces a complicated ground state which shows  phenomena
differed from that found in the coherent ground state?
A definite  answer to this question is a final goal of our study.
The first step to this goal is to examine extensively $n$-dependence 
of expectation values of operators for $\mid n \rangle $.

In the finite lattice system with  $\hat{H}_B $, an expectation value of an operator $\hat{O}$
is given by
$$ \langle E(N,B) \mid \hat{O} \mid E(N,B) \rangle 
= \sum_{n,n'} \langle n \mid \hat{O} \mid n' \rangle c_{n}(B) c_{n'}(B) \ .
$$
Here we notice that   $ \langle E(N,B) \mid \hat{O} \mid E(N,B) \rangle $ consists of  a
 $n$-independent term  $\langle \hat{O} \rangle $ and 
$n$-dependent terms  of $	\Delta \langle \hat{O} \rangle(n,n')$  of
$ \langle n \mid \hat{O} \mid n' \rangle $.
For the coherent states it is assumed implicitly that we observe only $\langle \hat{O} \rangle $.
For small $N$ systems or complicated macroscopic wave functions,  it is possible to observe 
$	\Delta \langle \hat{O} \rangle(n,n')$.
Therefore we would like to study on $ \langle n \mid \hat{O} \mid n' \rangle $.
These calculations have not been made yet, at least to my knowledge,
except the energy.
For the energy, previous works  have  confirmed the $n^2/N$ dependence numerically\cite{rev,MMcst1,MMcst2}.

 Our study is made on   the XY quantum antiferromagnet on a triangular lattice because of 
 followings reasons.
The first reason is that  this system has the U(1) symmetry, the  simplest one among continuous symmetry 
groups, which  are indispensable for existence of the NG boson.
Due to the same reason there has been many works on the  XY quantum antiferromagnet, specially the square lattice model
\cite{XY1,XY2}. They support that the XY model gives us essential properties of 
the Heisenberg model.

Another reason is that  previous works have found the excellent  trial state for the XY antiferromagnet on the triangle lattice
\cite{Singh,Capri}.
We can expect that the variational Monte Carlo(VMC) method is powerful in this system.
By the VMC method it becomes possible to study the NG boson in a large lattice,
which is  needed for study of excitation with small wave vectors.
Since the trial state is  essential, we must examine reliability of this state.
For this examination  we use the stochastic state selection(SSS) method\cite{MM1,MMss,MM2,MM3,MMtri,MMeq}.
Note that it is  difficult to calculate only by the VMC method.

 A plan of this paper is as follows. In section 2 we describe the model and  calculation methods.
In section 3 we present numerical results.
Lattice sizes are $N=48$ and $N=108$.  Calculations in $N=36$ and $N=324$ are made
for energy.
After fixing parameters of the trail state in the first subsection,  we show the lowest energy for the state with a fixed value of
the charge $\hat{Q} $ in subsection 3.2.
Also we employ the SSS method
to estimate quality of the trial state used in study by the VMC. In  subsection 3.3 we present
results for expectation values of a symmetry breaking operator.
In a next subsection we will show results of the structure functions,  which are
 determined by
the property of the NG boson\cite{Neuberger}.
In subsection 3.5 we will give a direct calculation of energy of excitation at a small wave vector.
By this calculation we obtain a velocity in this system\cite{velocity1,velocity2,velocity3}.
A final section is devoted to a summary and  discussions.
In  appendix A we make a brief description on the VMC method and the SSS method.
Here we describe a way to connect the SSS method with the VMC method.
Another appendix gives us a discussion on $n$-dependence of the expectation value of
the spin operator.

\section{ Model and calculation methods}
A system  which we study  is the quantum XY antiferromagnet of spin one-half  on the triangular lattice.
Its Hamiltonian is given by 
\begin{equation}
\hat{H} = \sum_{i,j} ( \hat{S}_x(i) \hat{S}_x(j) + \hat{S}_y(i) \hat{S}_y(j) )  \ .
\label{H}
\end{equation}
Here $ \hat{S}_x(i) $, $   \hat{S}_y(i) $, $  \hat{S}_z(i)  $ are   x-component, y-component and z-component
of one-half spin operators on the $i$-th site and the sum runs over all bonds of the $N$-site lattice.
This Hamiltonian has the U(1) symmetry, whose charge operator is defined by
\begin{equation}
 \hat{Q}= \sum_{i=1,\dots,N} \hat{S}_z(i) \ . 
\label{charge}
\end{equation}

Since we would like to make calculations in a large lattice,
we use the VMC method.
In the XY spin system on the triangular lattice,
previous works give us  the good trial state, $\mid \Psi \rangle $.
\begin{equation}
 \mid \Psi \rangle = \sum_{\{s_i\} } \mid \{s_i\} \rangle c(\{s_i\}) \ ,
\label{Psi}
\end{equation}
where $ \mid \{s_i\} \rangle $ is a basis state.
Here $ \{s_i\} =(s_1,s_2, \dots, s_N) $ , where  $ s_i=-1/2$, or $1/2$ .
The trial state \cite{Capri} is given by
\begin{equation}
    \mid \Psi \rangle = \sum_{(s_i)} Texp(g_1 \sum_{i,j} g_{i j} s_i s_j ) 
\mid \{s_i\}  \rangle \ ,
\label{trial}
\end{equation}
$$ T=exp( i\frac{2\pi}{3}\sum_{i \in B } s_i - i\frac{2\pi}{3}\sum_{i \in C } s_i )T_3  \ ,$$
$$ T_3=exp( i \beta \sum_{i.j.k} \gamma_{i j k} s_i s_j s_k ) \ , $$
$$ g_{i j}  = \sum_{\vec{k}} exp(i\vec{k}\cdot(\vec{x}_i - \vec{x}_j )) v(\vec{k})  \ ,$$
 $$v(\vec{k}) =1-\frac{1}{\sqrt{1-\gamma(\vec{k})}} \ , $$
$$ \gamma(\vec{k})=\{ cos(k_x)+2cos(k_x/2)cos(\sqrt{3}k_y/2) \}/3 \ ,$$
where we should 
note that each site is  categorized  into three sublattices,  A-sublattice, B-sublattice and C-sublattice.
Also $\gamma_{i j k}  s_i s_j s_k $ is a three body interaction, which is given  in Ref.\cite{Capri}.
Note that $ \mid \Psi \rangle $ is a sum of $ \mid n \rangle_\Psi$.
$$  \mid \Psi \rangle  = \sum_{n=-N, \dots , N}  \mid n \rangle_\Psi C_n \ . $$
Here $ \hat{Q} \mid n \rangle_\Psi= n  \mid n \rangle_\Psi$ .
This  trial state has two parameters, $ g_1$ and $ \beta $.
We search the minimum energy state of $n=0$ by changing these parameters.
After fixing values of $ g_1 $ and $ \beta$,
we calculate an expectation value of the Hamiltonian squared
in order to estimate quality of the trial state.
For these calculations we use the SSS method.
Then we will confirm that a difference between the expectation value of the Hamiltonian squared
and the square of the expectation value of the Hamiltonian  is small.

In the infinitely large lattice, properties of the NG boson come from a following
equation 
$$  \langle G \mid [ \hat{Q} , \hat{\phi} (i) ] \mid G \rangle =
 \langle G \mid \delta\hat{\phi} (i)  \mid G \rangle   \ ,
$$
using the ground state $ \mid G \rangle $.
%
If the right-hand expectation value is not zero, the SSB occurs.
However for a finite size system, we have only one state $ \mid n=0 \rangle$ for the lowest energy
and this state is the eigen state of $\hat{Q}$.
Therefore this expectation value vanish for a finite size lattice.
Instead of  the equation, we use a following equation for a finite size lattice.
$$  \langle m \mid [ \hat{Q} , \hat{\phi} (i)  ] \mid n \rangle = \langle m \mid \delta\hat{\phi} (i)  \mid n \rangle  \ .
$$
When  the right-hand side of the above is not be zero,
we see a signal of  the SSB in  a finite size lattice.
In our study  we adopt $\hat{S}_+(i)=\hat{S}_x(i) +i  \hat{S}_y(i)$
as $\hat{\phi} (i) $.
 The above equation becomes
\begin{equation}
 \langle n+1 \mid [ \hat{Q} , \hat{S}_+(i) ] \mid n \rangle = \langle n+1 \mid \hat{S}_+(i)  \mid n \rangle \ .
\label{exp-vl}
\end{equation}

This trial state $ \mid \Psi \rangle $ is constructed using an assumption 
that each site is  categorized  into three sublattices,  A-sublattice, B-sublattice and C-sublattice,
 where  directions of a unit vector are  $ 0$, $ 2\pi/3$ and $ 4\pi/3$ in $x-y$ plane
for these sublattices.
Therefore the expectation values of $\hat{S}_+(i)  $ for the $i$-th site  on a sublattice  differ from one on the another.
For getting the same value for every site, we make a rotation around the $z$-direction of spins.
For the $i$-th site on the A-sublattice we make no rotation.
$$   \hat{S}_x^R(i)=  \hat{S}_x(i)  \ ,  \ \ 
  \hat{S}_y^R(i)=  \hat{S}_y(i)  \ .$$
For the $i$-th site on the B-sublattice we make a rotation with an angle $2\pi/3$.
 $$   \hat{S}_x^R(i)= -\frac{1}{2}  \hat{S}_x(i) -\frac{\sqrt{3}}{2}\hat{S}_y(i) \ , \  \
 \hat{S}_y^R(i)= \frac{\sqrt{3}}{2}  \hat{S}_x(i) -\frac{1}{2}\hat{S}_y(i) \ .$$
For the $i$-th site on the C-sublattice we make a rotation with an angle $4\pi/3$.
 $$   \hat{S}_x^R(i)= -\frac{1}{2}  \hat{S}_x(i) +\frac{\sqrt{3}}{2}\hat{S}_y(i) \ , \  \
    \hat{S}_y^R(i)= -\frac{\sqrt{3}}{2}  \hat{S}_x(i) -\frac{1}{2}\hat{S}_y(i) \ .$$
 In this representation we have the same expectation value of spin operators, 
 $ \hat{S}_+^R(i) =\hat{S}_x^R(i)+ i \hat{S}_y^R(i)$   , being independent of the  site.
\begin{equation}
 \langle n+1 \mid \hat{S}_+^R(i) \mid n \rangle = v(n)  \ .
\label{exp-vl2}
\end{equation}
Note that this expectation value depends on $n$.

Also using these the rotated spin operators $ \hat{S}_x^R (i)$, $\hat{S}_y^R (i)$, $\hat{S}_z^R (i) $,
we calculate  structure functions.
For this purpose we introduce spin operators that depend on  wave vectors.
\begin{equation}
  \hat{S}_r^R(\vec{k}) = \frac{1}{\sqrt{N}} \sum_i exp(i \vec{k} \vec{r}_i ) \hat{S}_r^R(i)  \ .
\label{mom}
\end{equation}
Here $ \vec{k}$ is a wave vector, $ \vec{r}_i = n_1(i) \vec{e}_1 + n_2(i) \vec{e}_2 $ is a site vector, 
which is defined by integer numbers $n_1(i)$ and $n_2(i) $,  and $r=x,y,z$.  Also 
$\vec{e}_1=(1,0), \vec{e}_2=(1/2,\sqrt{3}/2) $ are unit vectors on the triangular lattice.
In future descriptions of $ \hat{S}_r^R(\vec{k}) $, we omit the superscript  $R$ in order to avoid complexity.

In order to make clear a relation between the  NG boson and the structure function,
 we define the NG boson. A  boson operator $\hat{\Phi} (i)$  is given by the
annihilation  $\hat{a}(\vec{k}) $ and creation operators  $\hat{a}^\dagger (\vec{k}) $.

\begin{equation}
  \hat{\Phi} (i)= \frac{1}{\sqrt{N} } \sum_{\vec{k} } \frac{1}{\sqrt{2E(\vec{k}) }}
  [ \ exp(i \vec{k} \vec{r}_i )\hat{a}(\vec{k})  + exp(-i \vec{k} \vec{r}_i )\hat{a}^\dagger(\vec{k}) \ ] .
\label{NG1}
\end{equation}
%
where $E(\vec{k}) $ is a energy, and 
in the infinitely large size lattice, 
 $ E(\vec{k})= c\mid \vec{k}  \mid $ for a small $\mid \vec{k}  \mid $.
As previous works show\cite{Neuberger}, the NG boson appears in the charge current $\hat{S}_z(i)$.
$$  \hat{S}_z(i)= f \partial_t \hat{\Phi} (i) + {\rm other \ terms } \ . $$
Here $\partial_t $ is a time derivative and we can neglect  $ {\rm other \ terms }$ 
 for a small wave vector. Also $ f$ is a constant.
From these discussions we have an equation in the infinitely large lattice.
\begin{equation}
  \langle G \mid \hat{S}_z(\vec{k})  \hat{S}_z(\vec{k}) \mid G \rangle = 
\frac{f^2   c \mid \vec{k} \mid }{2 }
  \ ,
\label{NG2}
\end{equation}
 for a small $\mid \vec{k} \mid $.
We will examine this equation carefully in the finite size lattice.

Also the field theoretical argument shows that
the NG-boson appears in the spin operator $  \hat{S}_y(i) $,
$$ \hat{S}_y(i) =  Z\hat{\Phi}(i) + {\rm other \ terms }  \ .$$
Therefore we have
\begin{equation}
  \langle G \mid \hat{S}_y(\vec{k}) \hat{S}_y(\vec{k}) \mid G \rangle = 
\frac{ Z^2}{2c \mid \vec{k} \mid } \ ,
\label{NG2}
\end{equation}
 for a small $\mid \vec{k} \mid $ and  the infinitely large size lattice.
We will study the above equation in subsection 3.4.

\section{ Results}
\subsection{ The trial state}
As described in section 2, 
the trail state is determined completely by parameters  $g_1$ and $\beta$.
By changing values of  $g_1$ and $\beta$, we find the minimum value
of the expectation of the Hamiltonian. By the minimum value
we obtain the best values of  these parameters.
As said previously, the trail state is given without fixing values of
$\hat{Q}$. That is
$$ \mid \Psi \rangle = \sum_{n=-N/2}^{N /2 } \mid n \rangle_\Psi C_n \ \ .$$
In a following discussion we omit a subscription "$\Psi$" of
a state $\mid n \rangle $ in order to avoid  complexity.
Here the state $\mid n \rangle $ is a complex state.
$$  \mid n \rangle =   \mid n \ ,\ R \rangle +\mid n \ , I \ \rangle i  \ \   . $$
For finding the minimum value of the expectation value of the Hamiltonian,
we use only a real state $ \mid 0 \ ,\ R \rangle $ for
avoiding statistical fluctuations due to the redundant complexity.
After finding  the minimum value,
we have confirmed a orthogonality of the real state and the imaginary state,
and the same value of the expectation value.  
$$   \langle 0\ ,  I  \mid 0 \ ,R \rangle  = 0  \ \ ,$$
$$ \langle 0\ , R   \mid \hat{H} \mid 0 \ , I  \rangle  = \langle 0\ , I  \mid \hat{H} \mid 0 \ , I  \rangle  \  \ .$$
The expectation value of the Hamiltonian is denoted by
$$ E(N,n)= \langle n\ , R \mid \hat{H} \mid n \ ,R \rangle  \ \ .$$
As results we obtain
$\beta=0.09 $, which is the same value  for all lattices with $N= 36,48,108,324$.
Obtained values for $g_1$ are given in  Table 1.
\begin{table}[t]
\begin{center}
\begin{tabular}{|c|c|c|c|c|}
\hline
$N $ & 36 & 48 & 108 &   324   \\  \hline
$g_1$  &  0.07 &  0.05 &  0.023 &  0.0075     \\  \hline
$ E(N,0) /N$  & $-0.40794 $  & $ -0.406718  $ & $ -0.405383$ & $  -0.404905  $ \\
  & $ \pm 0.1 \times 10^{-4} $  & $  \pm 0.8 \times 10^{-5}  $ &
 $ \pm 0.4\times 10^{-5} $ & $  \pm 0.15\times 10^{-4} $ \\
\hline
\end{tabular}
\end{center}
\caption{Values of parameter $g_1$ of the trial state and the minimum energy for the lattice size $N$.}
\label{table:1}
\end{table}
 If  $g_1$ is a function of the lattice size $N$,
we guess a function form,
 $$  g_1(N) \sim 0.25/N  \ \ . $$
\subsection{ Energy}
First we show results on the minimum expectation value of $ \hat{H} $ for  charge $n=0$ in lattices of  various  size $N$ ,
which are given in Table \ref{table:1}.
We would like to examine reliability of these results.
For $ N=36 $ we can compare it with the exact value, $ E_0/N=-0.41095 $ that is obtained by the
exact diagonarization\cite{Capri}. A difference between the exact value and the VMC value is about  $0.7 \%$

Next on  examinations for $N=48, 108$,
 we use the SSS method to calculate expectation values of
the Hamiltonian square $ \hat{H}^ 2 $.
By this calculation, we obtain a following ratio.
$$\delta(E)= ( \langle 0 \mid \hat{H}^2 \mid 0 \rangle - (\langle 0 \mid \hat{H} \mid 0 \rangle)^2 )
/(\langle 0\mid \hat{H} \mid 0 \rangle)^2 \ \ . $$
This ratio gives us rough estimations on differences between results and the exact values.
$$ \delta(E)   = 1.1\times 10^{-3} \pm 0.2\times 10^{-3} \ \ \ for \ N=36 \ \ ,
$$
$$ \delta(E) = 4.1\times 10^{-4} \pm 1.6\times 10^{-4 } \ \ \ for \ N=48  \ \ ,
$$
$$ \delta(E)= 4.2\times 10^{-4} \pm 0.4\times 10^{-4} \ \ \ for \ N=108  \ \ .
$$
For $N=36 $ the ratio $0.11 \% $ is somewhat smaller than the difference $0.7\% $.
Quantitative estimations on the difference from the exact value are difficult, but quite small values in 
these results   justify our  study on the SSB by the VMC.

From obtained energy we can have  size dependence of the energy, which is
$$ E(N,0)/N= e_0 +a/N^{3/2}\ \  ,$$
where
$e_0= -0.4048  \pm 0.0001$  and  $ a=-0.65 \pm 0.02 $. 
%
%

\begin{figure}[ht]
\begin{center}
\centering\includegraphics[width=2.5in]{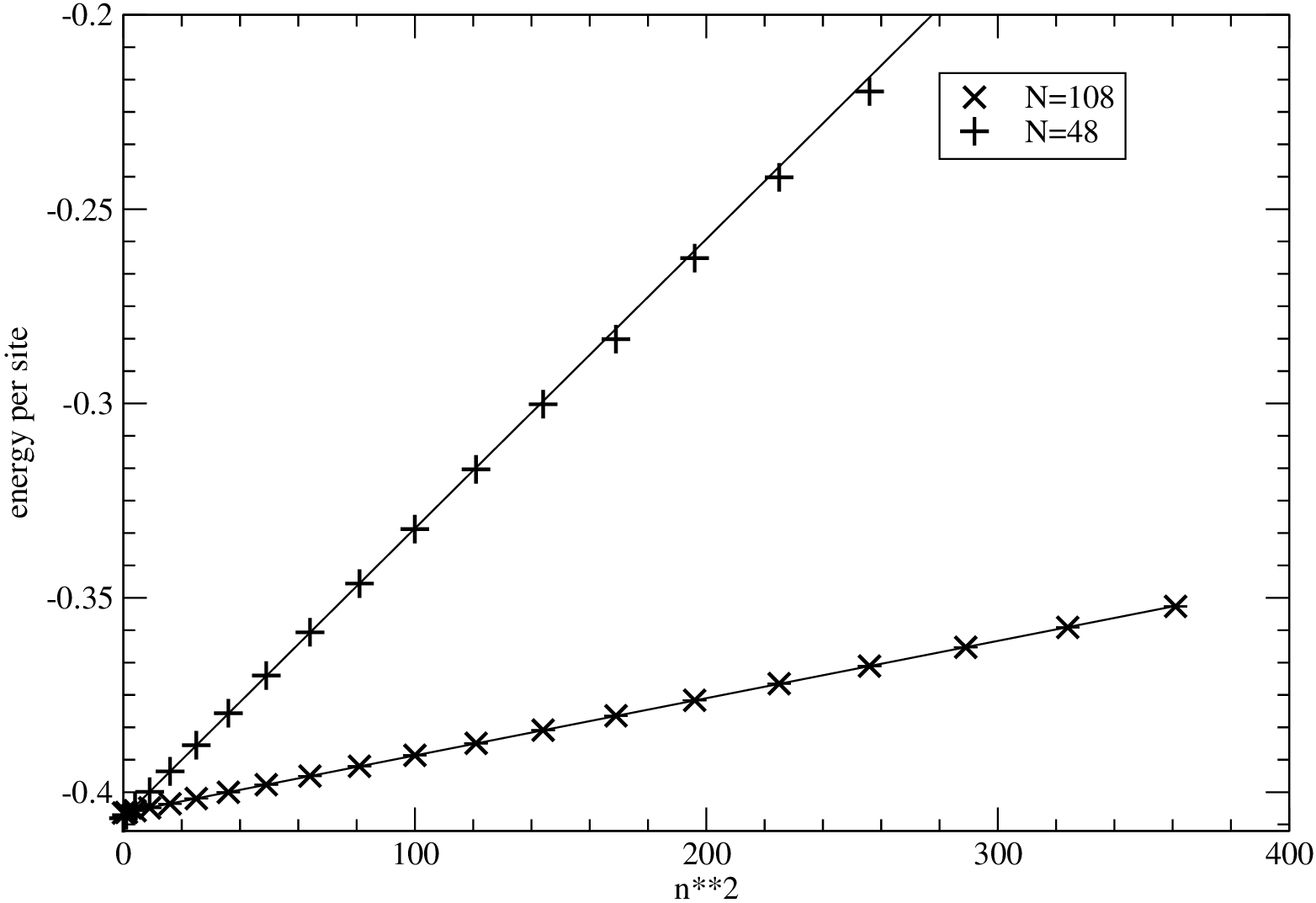}
\caption{ The lowest energy for each {\it n}.
Errors are small so that we do not plot  error bars.
The lines are given by applying the least square fit to data.
}
\label{fig:fig1}
\end{center}
\end{figure}

Next we will show results on expectation values of $\hat{H} $ for non-zero values $ n$.
Previous studies based on the SWT and numerical approaches 
show that dependence of the energy on $n$  is
\begin{equation}
E(N,n)/N = E(N,0)/N + b\ n^2/N^2 \ \ .
\label{E-dep}
\end{equation}
For $ N=48$ and $N=108 $ we plot $E(N,n)/N$ as $n^2 $ in Fig.\ref{fig:fig1}.
These results strongly support the above  dependence on $n$. Also we should note that
the above dependence (\ref{E-dep}) is acceptable for quite large $n$.
By the least square fitting,
they are $b=1.715 \pm 0.004$  for $N=48 $ and $b=1.726 \pm 0.002$  for $N=108 $.
%

\subsection { Expectation values of $\hat{S}_+(i) $}

As described in section 2,  an expectation value $v(n) $  of a operator  $ \hat{S}_+(i)  $, 
which show the breaking of  the $U(1)$ summery, 
is given by Eq.(\ref{exp-vl}).  
This equation assumes that these are independent of sites and real.
In order to verify these assumption,
we calculate a standard deviation on sites and imaginary parts of $v(n) $.
The site averages of real and imaginary parts of the expectation value for $ n=0$ are
\begin{eqnarray}
 & &\langle 1 \mid \hat{S}_+ (i) \mid 0  \rangle 
=   \nonumber \\
& &0.4815 \pm 0.0002   + i ( -1.27  \pm 1.40) \times 10^{-4} \ \ for \  N=48 \ \  ,
\nonumber
\end{eqnarray}
\begin{eqnarray}
  & & \langle 1 \mid \hat{S}_+ (i) \mid 0  \rangle 
=  \nonumber \\
 & &0.46425 \pm 0.000008   + i ( -1.11 \pm 1.52) \times 10^{-5}   \ 
 \ for \  N=108 \ \ .
\nonumber
\end{eqnarray}

From estimations  of the statics error, we can say that
the standard deviations on sites are $ 1.4\times 10^{-3}=2\times 10^{-4} \sqrt{N} $ for $N=48$,
and   $8.3\times 10^{-4} = 8\times 10^{-5} \sqrt{N} $ for $N=108$.
These values show that the above assumptions  are  justified numerically.
Dependence of  expectation value $v(n)$  on  $n$ is  shown
in Fig. \ref{fig:fig2}. Here the horizontal axis is  denoted by $n(n+1) $, and  a vertical axis is done by $ \{ v(n)\}^2 $.
We find that this  dependence is well described by a linear function of $ n(n+1) $.
$$
[ v(n)] ^2 = [v(0)]^2 + d \frac{n(n+1)}{N^2} \ \ .
$$
Values of $d$ are $ -0.82 \pm 0.02 $ for $N=48 $ and $-0.75 \pm 0.06$ for $ N=108$.
As discussed in Appendix B,  we have $d=-1$ if we assume that
we neglect contributions of the excited states
when we  calculate expectation values of $  [ \hat{S}_+(i) ,  \hat{S}_-(j) ]= 2 i \delta_{ij}  \hat{S}_z(j)  $.
This  small discrepancy on $d$ shows that we should not neglect these contributions.

\subsection{Structure functions }
In this subsection we show results on the structure function on a product of
$ \hat{S}_z(-\vec{k}) $ and $ \hat{S}_z(\vec{k}) $,
which is defined in a finite size system by
\begin{equation}
 F_{zz}(\vec{k},n) = \langle n \mid \hat{S}_z(-\vec{k}) \hat{S}_z(\vec{k}) \mid n \rangle \ \ .
\label{fzz1}
\end{equation}
For $ \mid \Psi \rangle $, we have
$$ \langle \Psi \mid \hat{S}_z(-\vec{k}) \hat{S}_z(\vec{k}) \mid \Psi  \rangle 
 = \sum_n  \langle n \mid \hat{S}_z(-\vec{k}) \hat{S}_z(\vec{k}) \mid n  \rangle C_n^* C_n$$
$$ =\sum_n F_{zz}(\vec{k},n) \mid  C_n \mid^2   \ \ .$$
If  the NG boson dominates in the structure function for small
wave vectors,
$$ F_{zz}(\vec{k},n) = \frac{f^2}{2}  c\mid \vec{k} \mid +{\rm n-dependent\  terms} \ \ . $$
To check this form,
we plot results on $F_{zz}(\vec{k},n) $ as a function of
$  \mid \vec{k} \mid  $ in Figures \ref{fig:fig3}, \ref{fig:fig4}.
For $\mid \vec{k} \mid  \leq 2$, we can see a linearity of   $\mid \vec{k} \mid  $
for $N=48 $ and $N=108 $.
This linearity strongly supports that the NG boson dominates in the low energy spectrum.
Also we find that  $ F_{zz}(\vec{k},n)$  depends on $n$.
In order to see this  $n$-dependence in detail,
we plot $ F_{zz}(\vec{k},n)$ as a function of $n^2 $.
From  these plots  we find that $F_{zz}(\vec{k},n) $ are linear functions  of $n^2$,
which are
$$   F_{zz}(\vec{k},n)  = F_{zz}(\vec{k},0 ) +g_{zz}(\vec{k})n^2\ \ .$$
Next we plot $ g_{zz}(\vec{k}) $ as a function of $ \mid \vec{k} \mid $.
By results we guess that this is a linear function of $ \mid \vec{k} \mid $.
We make the least square fitting on this function using data of $ \mid \vec{k} \mid \leq 2$.
\begin{equation}
     F_{zz}(\vec{k},n)= \frac{f^2}{2} c \mid \vec{k} \mid +(\alpha_{zz}  + \beta_{zz} \mid \vec{k} \mid )n^2/N^2 \ \ .
\label{fzz2}
\end{equation}
From these  figures we obtain for $N=48$ and $N=108$,
\begin{eqnarray}
  \frac{f^2}{2} c & =& 0.130 \pm 0.004\ , \ \  \alpha_{zz} = 0.26 \pm 0.07\ ,  
\nonumber \\
\   \beta_{zz}&= &-0.60 \pm 0.06
 \ \ \  {\rm  for } \  \ N=48 \ \ , \nonumber 
\end{eqnarray}
\begin{eqnarray}
    \frac{f^2}{2} c  & =&   0.126 \pm 0.003\ , \ \  \alpha_{zz} =0.45 \pm 0.07\ ,  
\nonumber \\
\ \   \beta_{zz}   &= & -0.67 \pm 0.08
\ \ \ {\rm  for } \  \ N=108\ \ .  \nonumber 
\end{eqnarray}
%
%
  A squared error of the expression (\ref{fzz2}) is calculated by an average of square of differences between
data and the expression. The error par one data is
$ 0.0036 $ for $N=48$,  and $0.0025$ for  $N=108$.
These small values imply that the function (\ref{fzz2}) can describe  quite well the data.

Next we show results on the structure function on a product of
$ \hat{S}_y(-\vec{k}) $ and $ \hat{S}_y(\vec{k}) $,
which is defined in a finite size system by
\begin{eqnarray}
& &  F_{yy}(\vec{k}, n)  = -\langle n+1 \mid \hat{S}_+(-\vec{k)}  \hat{S}_+(\vec{k}) \mid n-1\rangle /2
\nonumber \\
                        &   -  &\langle n-1 \mid \hat{S}_-(-\vec{k})  \hat{S}_-(\vec{k}) \mid n+1\rangle /2 
\nonumber \\
                      &   + &\langle n \mid \hat{S}_-(-\vec{k})  \hat{S}_+(\vec{k}) \mid n\rangle /2
\nonumber \\
                         & + &\langle n \mid \hat{S}_+(-\vec{k})  \hat{S}_-(\vec{k}) \mid n\rangle  /2  \ \ .
\label{fyy1}
\end{eqnarray}
For $ \mid \Psi \rangle $, we have
\begin{eqnarray}
 &  & \langle \Psi \mid \hat{S}_y(-\vec{k}) \hat{S}_y(\vec{k}) \mid \Psi  \rangle =
\nonumber \\
   &  & \sum_n \{  -\langle n+1 \mid \hat{S}_+(-\vec{k)}  \hat{S}_+(\vec{k}) \mid n-1\rangle /2 C^*_{n+1}C_{n-1}
\nonumber \\
     &- &                \langle n-1 \mid \hat{S}_-(-\vec{k})  \hat{S}_-(\vec{k}) \mid n+1\rangle /2  C^*_{n-1}C_{n+1}
\nonumber \\
                      &   + &\langle n \mid \hat{S}_-(-\vec{k})  \hat{S}_+(\vec{k}) \mid n\rangle /2  C^*_{n-1}C_{n+1}
\nonumber \\
                        &  + &\langle n \mid \hat{S}_+(-\vec{k})  \hat{S}_-(\vec{k}) \mid n\rangle  /2   C^*_{n}C_{n}  \}
\nonumber \\
& = &  \sum_n F_{yy}(\vec{k}, n)  C^*_{n}C_{n}
\nonumber \\
&-  &  \sum_n \{  \langle n+1 \mid \hat{S}_+(-\vec{k)}  \hat{S}_+(\vec{k}) \mid n-1\rangle /2 
     ( C^*_{n+1}C_{n-1} - C^*_{n}C_{n} )
\nonumber \\
    & +&      \langle n-1 \mid \hat{S}_-(-\vec{k})  \hat{S}_-(\vec{k}) \mid n+1\rangle /2  
  ( C^*_{n+1}C_{n-1} - C^*_{n}C_{n}   )   \} \ \ .
\label{fyyG}
\end{eqnarray}
This equation shows that if we know coefficients $C_n$,  we calculate the expectation of
$ \langle \Psi \mid \hat{S}_y(-\vec{k}) \hat{S}_y(\vec{k}) \mid \Psi  \rangle $.

If contributions form the NG boson dominate in the structure function for small
wave vectors,
$$ F_{yy}(\vec{k},n) = \frac{Z^2}{2 c\mid \vec{k} \mid } +n-{\rm dependent\  terms} \ \ .$$
To check this form,
we plot results on $F_{yy}(\vec{k},n) $ as a function of
$  1/ \mid \vec{k} \mid  $ in Figures \ref{fig:fig5}, \ref{fig:fig6}.
For $\mid \vec{k} \mid  \leq 2$, we can see a linearity of   $ 1/ \mid \vec{k} \mid  $
for $N=48 $ and $N=108 $.
This linearity strongly supports that the NG boson dominates in the low energy spectrum.
In order to see this  $n$- dependence,
we make the same discussion as that for $ F_{zz}(\vec{k},n)$.
First we plot $ F_{yy}(\vec{k},n)$ as a function of $n^2 $. Then
from  these plots  we find that $F_{yy}(\vec{k},n) $ is a linear function  of $n^2$,
which are
$$   F_{yy}(\vec{k},n)  = F_{yy}(\vec{k},0 ) +g_{yy}(\vec{k})n^2 \ \ .$$
Next we plot $ g_{yy}(\vec{k}) $ as a function of $ 1/ \mid \vec{k} \mid $.
Results suggest that this function is a linear function of $ 1/\mid \vec{k} \mid $.
If we make the least square fitting on this function,
we obtain
\begin{equation}
     F_{yy}(\vec{k},n)=\frac{ Z^2}{ 2 c \mid \vec{k} \mid } +(\alpha_{yy}  + \beta_{yy}/ \mid \vec{k} \mid )n^2/N^2\ \ .
\label{fyy2}
\end{equation}
From figures we obtain for $N=48$ and $N=108$,

\begin{eqnarray}
  Z^2/2 c    &=& 0.458 \pm 0.004   , \  \alpha_{yy} =0.90\pm 0.04  ,\  \nonumber \\
 \beta_{yy}   &=& -1.66 \pm 0.06\ \ \ {\rm  for } \   N=48 \  .\nonumber 
\end{eqnarray}
\begin{eqnarray}
 Z^2/2 c    &=&  0.439 \pm 0.005  , \  \alpha_{yy} =0.89\pm 0.12 ,\   \nonumber \\
\beta_{yy}  & &=-1.65 \pm 0.15\ \ \ {\rm  for } \   N=108\  .  \nonumber 
\end{eqnarray}
A squared error of the expression (\ref{fyy2}) is calculated by an average of square of differences between
data and the expression. The error par one data is
$ 0.0027$ for $N=48$,  and $0.0037$ for  $N=108$.
By these small values  it is justified that  the function (\ref{fyy2})  describes  the  data as well as that for $F_{zz}(\vec{k},n)$.

We make comments on a structure function of a product of
$ \hat{S}_y(-\vec{k}) $ and $\hat{S}_z(\vec{k}) $.
In descriptions by the NG boson  we have

$$  \langle \Psi  \mid \hat{S}_y(-\vec{k}) \hat{S}_z(\vec{k}) \mid \Psi \rangle 
= i Zf/2\ \ .
$$
Also from a commutation of 
$$ [ \hat{S}_y(\vec{k}_1),\hat{S}_z(\vec{k}_2)] =\frac{i}{ \sqrt{N} } \hat{S}_x(\vec{k}_1+\vec{k}_2)  , $$
we have $  Zf =v  $. In order to examine this equation, we calculate the structure 
function in the finite size lattice, which is defined by
\begin{eqnarray}
 F_{yz}(\vec{k}, n) &=& \frac{i}{ \sqrt{2} }\langle n+1 \mid \hat{S}_+(\vec{k})  \hat{S}_z(-\vec{k}) \mid n\rangle 
\nonumber \\
               &  - &   \frac{i}{ \sqrt{2} } \langle n \mid \hat{S}_-(\vec{k}) \hat{S}_z(-\vec{k}) \mid n+1\rangle \ \ .
\nonumber 
\end{eqnarray}
We  have  confirmed  numerically  that $F_{yz}(\vec{k}, n)$'s are independent of wave vectors $\vec{k} $ and their values agree
with $v(n)$ within error, although results are omitted. 

\subsection{ Energy for excitations  with small wave vectors}

In  calculations of the structure functions,
we do not see the  energy $\omega(\vec{k})$ of the excitation, so that
we could not determine a value of the velocity $ c$, which is defined by
$ c=  \omega(\vec{k})/ \mid \vec{k} \mid $. 
In order to obtain the velocity, we calculate the energy by a following method.
If  $ \mid \vec{k} ,n \rangle $ is a state of one NG boson with a wave vector $ \vec{k} $ and a charge  $n$,
we have $ \hat{H} \mid \vec{k} ,n \rangle  = \mid \vec{k} ,n \rangle \{ \omega(\vec{k}) + E(N,n)  \}$.
Note that $ E(N,n) $ is the lowest energy for the charge $n$.
It is difficult to get an eigenvalue so that we are contented with an expectation value of the $\hat{H} $,
which is given by
$$
 \omega_A(\vec{k}) = \langle \vec{k} ,n \mid \hat{H} \mid \vec{k} ,n \rangle - E(N,n)  \ \ .
$$
The results on  the structure function imply that the spin operator $\hat{S}_z(\vec{k}) $
can create a state of one NG-boson with a wave vector $ \vec{k} $ from the lowest energy state
$ \mid n \rangle $. Therefore we can approximate the state of one NG boson by
the lowest energy state that is operated by $\hat{S}_z(\vec{k}) $.
$$ \hat{S}_z(\vec{k}) \mid n \rangle \sim \mid \vec{k} ,n \rangle A  \ . $$
Here $A$ is a constant.
If we use this approximation, a calculated energy $ \omega_A(\vec{k}) $ is 
\begin{eqnarray}
 & \omega_A(\vec{k})   &  = \langle \vec{k} ,n \mid \hat{H} \mid \vec{k} ,n \rangle  -E(n) \ \ ,
\nonumber \\
& = &A^2\langle \vec{k} ,n \mid \hat{H} \mid \vec{k} ,n \rangle /(  A^2\langle \vec{k} ,n  \mid \vec{k} ,n \rangle )
-E(n) \ \ ,
\nonumber \\
& \sim &
\langle n \mid \hat{S}_z(-\vec{k}) \hat{H} \hat{S}_z(\vec{k}) \mid n \rangle/
 ( \langle n \mid \hat{S}_z(-\vec{k}) \hat{S}_z(\vec{k}) \mid n \rangle)  \nonumber \\
& -  &E(n) \ \ .
\label{NG_er}
\end{eqnarray}
Results of calculations of $\omega_A(\vec{k})$ are collected in Table 2.
\begin{table}[t]
\begin{center}
\begin{tabular}{|c|c|c|c|c|}
\hline
N & n & $(k_x,k_y)$  & $\omega$  & c \\  \hline
48 & 0 &  $(\pi/3,0)$   & $ 0.83 \pm 0.04 $  &  $0.79\pm 0.04$   \\  \hline
48 & 0 &  $(\pi/2,-\frac{\pi}{2 \sqrt{3}})$   & $ 1.46 \pm 0.03 $  &  $0.80\pm 0.03$   \\  \hline
48 & 6 &  $(\pi/3,0)$   & $ 0.85\pm 0.02 $  &  $0.81 \pm 0.02$   \\  \hline
48 & 6 &  $(\pi/2,-\frac{\pi}{2 \sqrt{3}})$   & $ 1.36 \pm 0.03 $  &  $0.75 \pm 0.02$   \\  \hline
48 & 10 &  $(\pi/3,0)$   & $ 0.69 \pm 0.04 $  &  $0.69 \pm 0.04$   \\  \hline
48 & 10 &  $(\pi/2,-\frac{\pi}{2 \sqrt{3}})$   & $ 1.33 \pm 0.03 $  &  $0.73 \pm 0.03$   \\  \hline \hline
108 & 0 &  $(\frac{2\pi}{9}, 0) $   & $ 0.56 \pm 0.03 $  &  $0.80\pm 0.04$   \\  \hline
108 & 0 &  $(\frac{\pi}{3}, -\frac{\pi}{3\sqrt{3}})$   & $ 1.01 \pm 0.02 $  &  $0.83\pm 0.02$   \\  \hline
108 & 0 &  $(\frac{4\pi}{9}, 0)$   & $ 1.16 \pm 0.02 $  &  $0.83\pm 0.02$   \\  \hline
108 & 0 &  $(\frac{5\pi}{9}, -\frac{\pi}{3\sqrt{3}})$   & $1.40 \pm 0.02 $  &  $0.76\pm 0.02$   \\  \hline
108 & 6 &  $(\frac{2\pi}{9}, 0) $   & $ 0.58 \pm 0.04 $  &  $0.83 \pm 0.06$   \\  \hline
108 & 6 &  $(\frac{\pi}{3}, -\frac{\pi}{3\sqrt{3}})$   & $ 1.02 \pm 0.03 $  &  $0.84\pm 0.03$   \\  \hline
108 & 6 &  $(\frac{4\pi}{9}, 0)$   & $ 1.11 \pm 0.03 $  &  $0.80\pm 0.03$   \\  \hline
108 & 6 &  $(\frac{5\pi}{9}, -\frac{\pi}{3\sqrt{3}})$   & $  1.42 \pm 0.03 $  &  $0.77\pm 0.02$   \\  \hline
108 & 10 &  $(\frac{2\pi}{9}, 0) $   & $ 0.53 \pm 0.03 $  &  $0.76 \pm 0.05$   \\  \hline
108 & 10 &  $(\frac{\pi}{3}, -\frac{\pi}{3\sqrt{3}})$   & $ 0.98 \pm 0.03 $  &  $0.81\pm 0.03$   \\  \hline
108 & 10 &  $(\frac{4\pi}{9}, 0)$   & $ 1.08 \pm 0.03 $  &  $0.77\pm 0.03$   \\  \hline
108 & 10 &  $(\frac{5\pi}{9}, -\frac{\pi}{3\sqrt{3}})$   & $  1.44 \pm 0.03 $  &  $0.78\pm 0.03$   \\  \hline
108 & 16 &  $(\frac{2\pi}{9}, 0)$   & $ 0.52\pm 0.2 $  &  $0.74\pm 0.04$   \\  \hline
108 & 16 &  $(\frac{\pi}{3}, -\frac{\pi}{3\sqrt{3}})$   & $ 0.94\pm 0.03 $  &  $0.78 \pm 0.03$   \\  \hline
\end{tabular}
\end{center}
\caption{energy of an excited state with a wave vector }
\label{table:2}
\end{table}
For $n=0$, the velocity is a constant for the small wave vector within
some error.
This value is a little larger than 0.75 that is calculated by the  linear spin wave theory.
On the $n$-dependence of the velocity we could not make a definite conclusion due to the large error.
Here we would like stress that these results are quite  non-trivial because  calculations on the
Hamiltonian are completely independent from those on the structure functions.

\section{ Summary and discussions}

Recent experiments on condensed matters as alkali atoms and superconductors require more complicated
 macroscopic wave functions, which differ essentially from the simple coherent state.
In order to understand phenomena by these wave functions in  quantum spin systems,
we have made theoretical study on states of  the SSB in these systems.
In this work  we used the VMC method for   numerical study.
In this method we have to assume that
the trial state is close to an exact eigen state of the Hamiltonian.
Based on this  assumption a study  was made for the XY quantum antiferromagnet
on the  triangular lattice, where the good trail state has been known.
Also in order to justify this assumption, we employed the SSS method to calculate
the square of the Hamiltonian. 
By results on the square we confirmed that the trial state was of a high quality.

Our numerical examination has been made   on  states that become  degenerate in a infinitely large lattice,
after the confirmation on dominance of a NG boson in low energy excitation. 
In the examination  we calculated  the energy,   the expectation value of the spin operator
and the structure functions of spin by fixing
 a  quantum number, which is a value of the  $z$ component of all spins.

Our results on numerical calculation in lattice sizes of 48 and 108 showed that
the energy of a small wave vector was  linear to a magnitude of the wave vector. In addition it was shown that
the expectation values  of the operators studied here were linear  functions of a square of the quantum number, $n^2$.
Though in our study this conclusion is limited to the XY model on the triangular lattice, we suppose that
the same conclusion can be obtained for other quantum spin models because
$n$-dependent terms are restricted partially by algebraic structures as seen in appendix B.

Final comments are made on experimental observations of the $n$-dependent terms.
Our  results show that for a ground state with $n \ll N $ and a large $N$, 
one can neglect contributions of $n$-dependent terms.
However if a external interaction is the charge operator $\hat{B}=h \hat{Q} $, 
the ground state is $\mid n* \rangle$, where $n*=h/2d \ N$. For this state there exists
a possibility of observing the $n$-dependent terms.
More detailed discussion will be made in future works.

\appendix
\section{ Appendix A}
In this appendix we discuss on  the VMC method and the SSS method.
A trial state $\mid \Psi \rangle $ is given by the coefficients  $c(\{s_i\}) $
on the basis state, as described in section 2.
$$   \mid \Psi \rangle = \sum_{\{s_i\} } \mid \{s_i\} \rangle c(\{s_i\})  \ \ .$$
In the VMC method, a probability variable $X$ is a basis state of $\mid \{ s_i \} \rangle $.
A probability  $ P_{VMC}(X=\{s_i\} ) $  for this variable  is defined    by 
$$    P_{VMC}(X=\{s_i\} ) = \mid c(\{s_i\} )\mid  / [ \sum_{\{s_j\}}  \mid c(\{s_j\} )\mid  ] \ \ .$$
In a Markov chain Monte Carlo method, what we need is only a ratio of probabilities, 
$ \mid c(\{s_i'\} )\mid  / \mid c(\{s_i\}  )\mid  $ so that we do not need the normalization factor, 
which corresponds to the denominator.
In this method  a expectation value of a operator $\hat{O} $ in the trial state $\mid \Psi \rangle $
is given by
\begin{eqnarray}
& & \langle \Psi \mid \hat{O} \mid \Psi \rangle / \langle \Psi \mid \Psi \rangle
\nonumber \\
& = &\frac{ \sum_{ \{s'_i\} } \sum _{ \{s_j\} }c^\dagger (\{s'_i\} )  O_{ \{s'_i\}, \{s_j\} } c(\{s_j\} )   }
{  \sum _{ \{s_i\} }  \mid c(\{s_i\} )  \mid^2    }  \ \ ,
\nonumber \\
& = & [ \ \sum_{ \{s'_i\} } \sum _{ \{s_j\} }c^\dagger (\{s'_i\} )  O_{ \{s'_i\}, \{s_j\}  }c(\{s_j\} )/ \mid  c(\{s_j\} ) \mid 
\nonumber \\
&\cdot &P_{VMC}( X=\{s_j\} ) \  ]
\nonumber \\
& / & [ \ \sum _{ \{s_i\} }  \mid c(\{s_i\} )  \mid     P_{VMC}( X=\{s_i\} )  \ ]  \ \ ,
 \nonumber 
\end{eqnarray}
where
$$   O_{ \{s'_i\}, \{s_j\}  }= \langle \{s' _i\} \mid \hat{O} \mid \{s_j\}  \rangle  \ \ .
$$
Here note that we do not need $ \langle \Psi \mid \Psi \rangle =1  $ .
Actually  the trial state we used in this study is not normalized.

While in the SSS method we give 
a probability variable  $X_{\{s_i\} } $ to each basis state $\mid \{s_i\} \rangle $,  whose value is $0$ or $a_{ \{s_i\}}$.
Here $1/ a_{ \{s_i\}}  =min(1,\mid c(\{s_i\} ) \mid  C ) \le 1$.  A probability 
$ P_{SSS}( X_{\{s_i\} } =  a_{ \{s_i\}}   ) = 1/a_{ \{s_i\}}$ , while  $P_{SSS}( X_{\{s_i\}}=0) =1 - 1/a_{ \{s_i\}}  $.  
An average of $ X_{\{s_i\} }$ is 1.
We multiply this probability variable $ X_{ \{s_i\} }$ to
a coefficient $  c(\{s_i\} ) $ and we have $ X_{\{s_i\} }c(\{s_i\} ) $, that results in 0 or $  phase( c(\{s_i\} ) )/C$.
Here  $ phase( x ) = x / \mid x \mid $. 
A state with a nonzero coefficient is called as a sampled state after the multiplication.
A value of $C$  determines  an average number of  sampled states. That is,
as  $C$ becomes  large, this number increases and the statistical error decreases.

We will describe a calculation method for an expectation value of the Hamiltonian squared, 
$ \langle \Psi \mid \hat{H}^2 \mid \Psi \rangle $.
 First by the VMC method  we  make sampling to collect $N_{smpl} $ basis states.
Next we operate the Hamiltonian $ \hat{H} $ to these sampled states.
After the operation,  a number  $N_{H} $ of the basis states with  nonzero coefficients
 is a few hundred times of $N_{smpl} $.
If we operate $ \hat{H} $ to $N_{H} $ basis states,  we need very huge number of basis states for 
calculations of its inner product with the trial state $\mid \Psi  \rangle $.
This huge number  makes a calculation difficult due to limited resources of  CPU time and computer memory.
In order to overcome this difficulty,  we employ the SSS method to $N_{H} $ basis states to 
reduce a number of basis states.
After a reduction of the basis states, we operate $ \hat{H} $ and 
calculate its inner product with the trial state $\mid \Psi  \rangle $.
By repeating these calculations, we make a statistical average for 
$ \langle \Psi \mid \hat{H}^2 \mid \Psi \rangle $.

\section{ Appendix B}
In this appendix we will present a discussion on  $n$ dependence of 
$ v(n) = \langle n+1 \mid \hat{S}_+(i) \mid n \rangle $.
We start  from   a commutation relation on the spin operator.
$$   [ \hat{S}_+(i) , \hat{S}_-(j) ] = 2 \delta_{ij}  \hat{S}_z (i)  \ \ .$$
By the state  $ \mid  n \rangle $  we calculate 
a expectation value of operators in both sides.
$$ \langle n \mid  [ \hat{S}_+(i) , \hat{S}_-(j) ] \mid  n \rangle = 
\delta_{ij}  \langle n \mid  \hat{S}_z (i)  \mid  n \rangle \ \ , $$
$$ \langle n \mid   \hat{S}_+(i) \hat{S}_-(j)  \mid  n \rangle
 = \langle n \mid   \hat{S}_+(i)   \mid  n-1 \rangle  \langle n-1 \mid   \hat{S}_-(j)  \mid  n \rangle  \ \ 
$$
$$
+\sum_{e} \langle n \mid   \hat{S}_+(i)   \mid  n-1,e \rangle  \langle n-1,e \mid   \hat{S}_-(j)  \mid  n \rangle\ \ .
$$
Here   $ \mid  n-1,e \rangle $ denotes an  excited state of the charge of  $n-1$.
We assume that the second term can be neglected as described  in Ref.\cite{Neuberger}.
Noting that
$ \langle n-1 \mid   \hat{S}_-(j)  \mid  n \rangle= \langle n \mid   \hat{S}_+(j)  \mid  n-1\rangle ^\dagger
= v(n-1)^\dagger $,
we obtain a following relation by making a sum over  $i,j$ 

$$  \{ v(n-1) \}^2 -  \{ v(n) \}^2 =2n/N^2\ \ .$$
Therefore we have
\begin{equation}
   \{ v(n) \}^2= \{ v(0) \}^2 - n(n+1)/N^2 \ \ .
\label{vvv}
\end{equation}
Here note that $ v(n) $ is real as explained in section 3.3.

\begin{figure}[!h]
\begin{center}
\centering\includegraphics[width=2.5in]{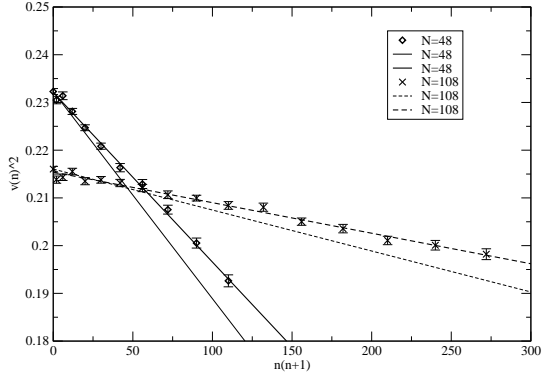}
\caption{ Squares of the expectation values  for each {\it n}.
The bold lines are given by applying the least square fit to data.
The slim lines are given by equation (\ref{vvv}).
}
\label{fig:fig2}
\end{center}
\end{figure}

\begin{figure}[!h]
\begin{center}
\centering\includegraphics[width=2.5in]{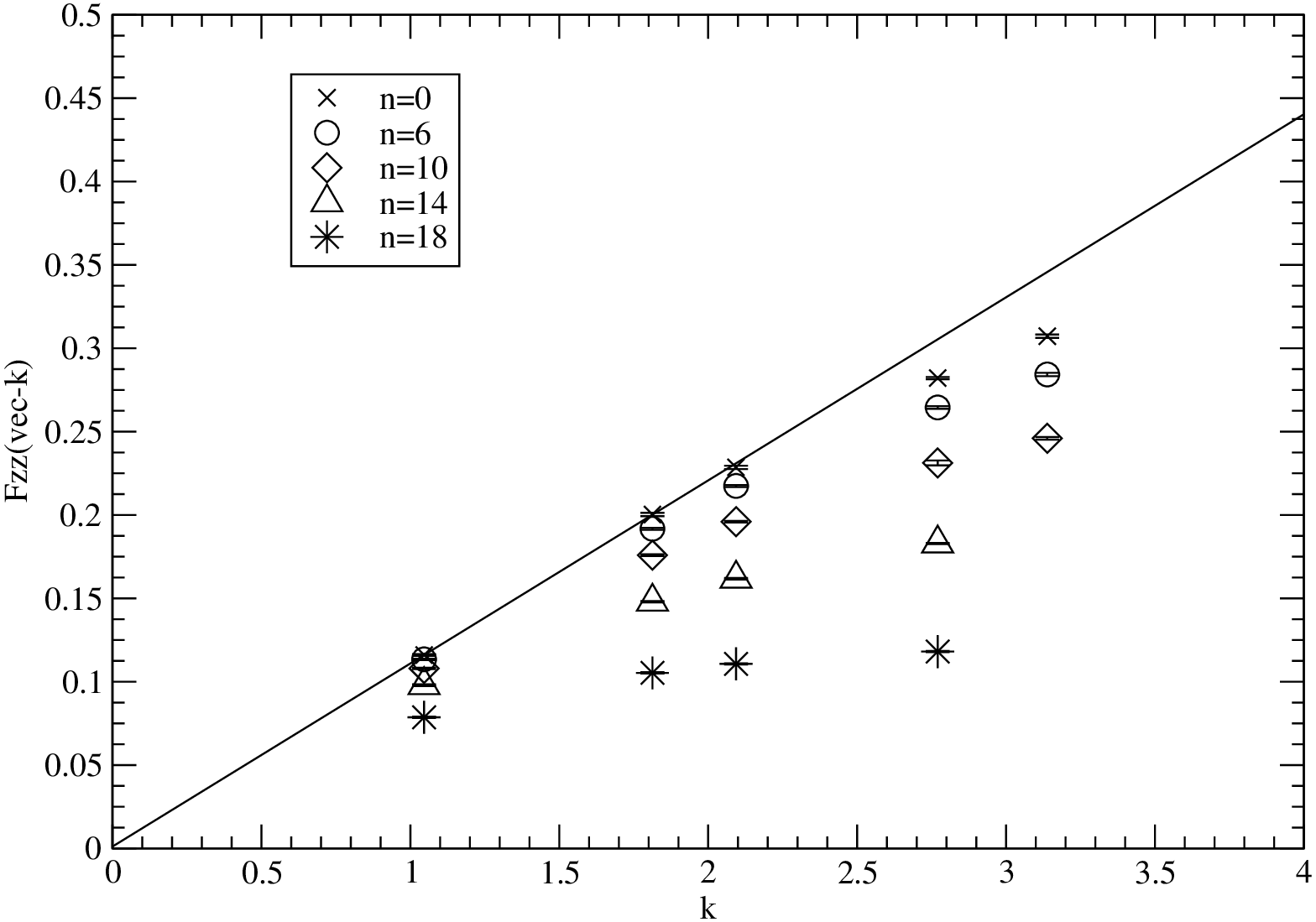}
\caption{ Structure Function $F_{zz}(\vec{k}) $  for $N= 48$.
The line is given by applying the least square fit to data of $n=0$, for which the magnitude of wave vectors  is less than $2$.
}
\label{fig:fig3}
\end{center}
\end{figure}

\begin{figure}[!h]
\begin{center}
\centering\includegraphics[width=2.5in]{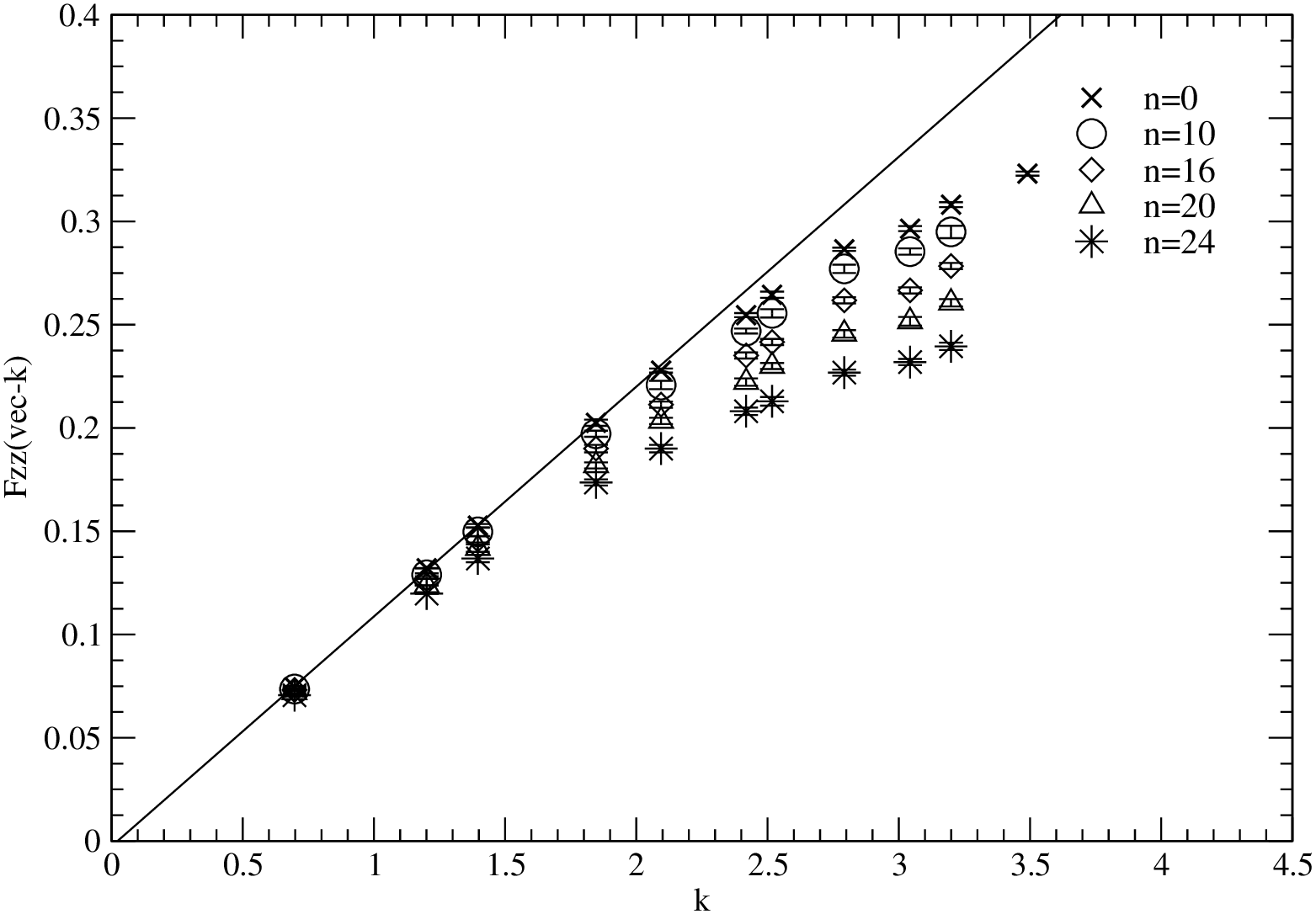}
\caption{ Structure Function $F_{zz}(\vec{k})$ for $N=108$.
The line is  given by applying the least square fit to data of $n=0$,
 for which the magnitude of wave vectors  is less than $2$.
}
\label{fig:fig4}
\end{center}
\end{figure}

\begin{figure}[!h]
\begin{center}
\centering\includegraphics[width=2.5in]{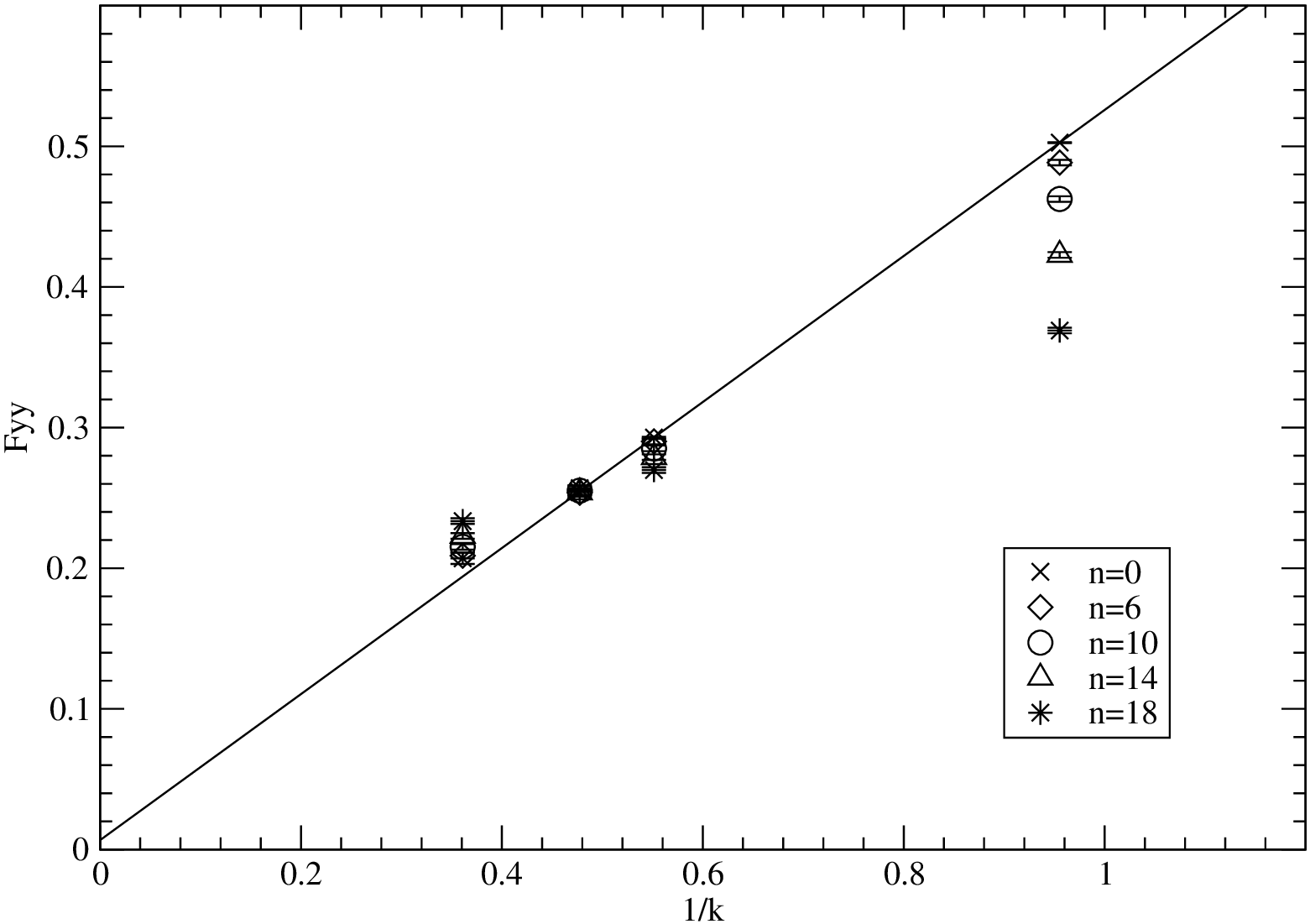}
\caption{ Structure Function $F_{yy}(\vec{k}) $ for $N=48$.
The line is given by applying the least square fit to data of $ n=0$,
 for which the magnitude of wave vectors  is less than $2$.
}
\label{fig:fig5}
\end{center}
\end{figure}

\begin{figure}[!h]
\begin{center}
\centering\includegraphics[width=2.5in]{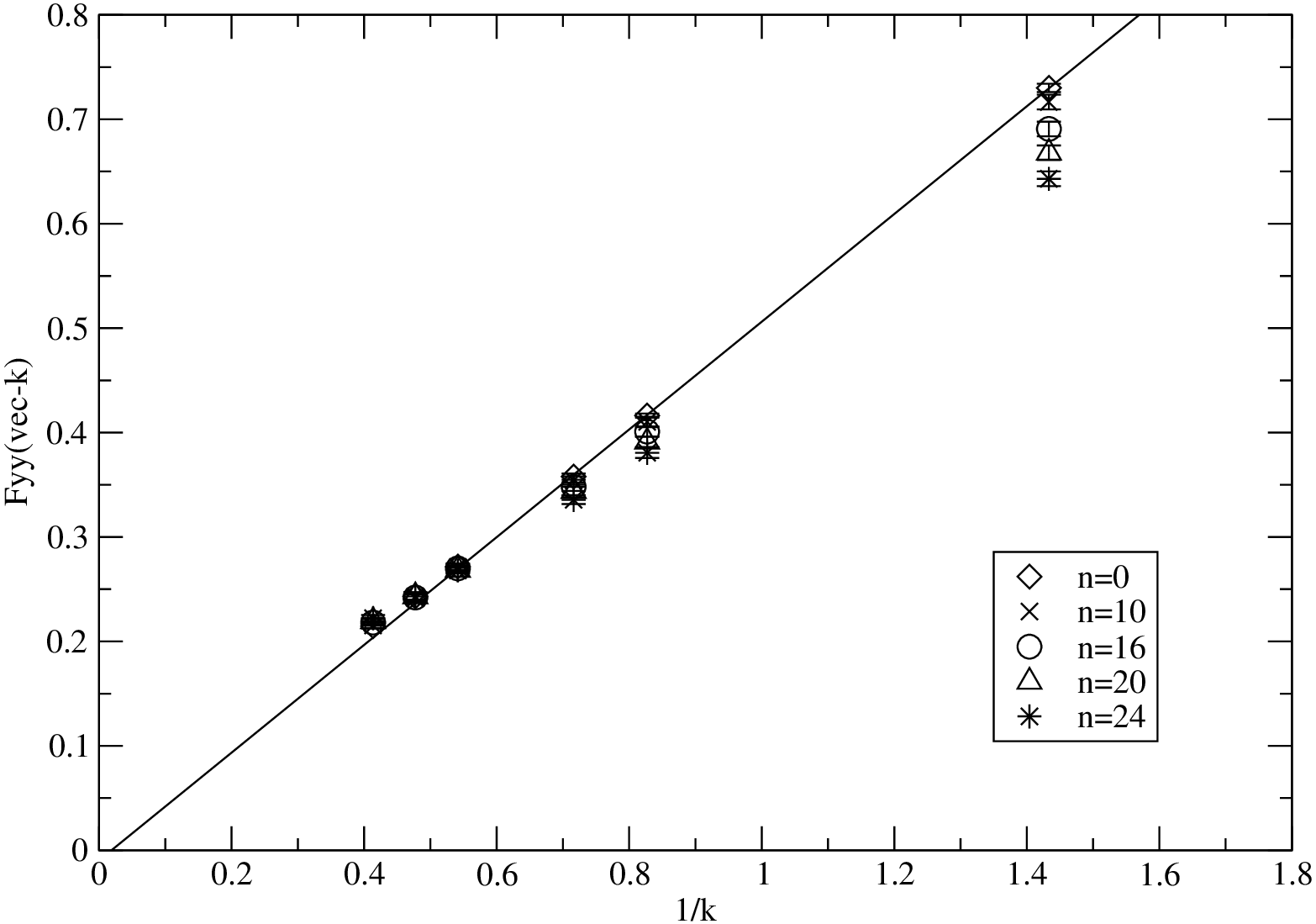}
\caption{ Structure Function $F_{yy}(\vec{k}) $ for $N=48$.
The line is given by applying the least square fit to data of $n=0$,
for which the magnitude of wave vectors  is less than $2$.
}
\label{fig:fig6}
\end{center}
\end{figure}

The above discussion  explains well the linear dependence of $\{ v(n)\} ^2$ on $n(n+1) $ which is found in Fig. \ref{fig:fig2}, 
although decreasing rates differ from the above estimations even for $N=108$.
It may imply that it is too rough to neglect contributions from excited states in calculating 
$ \langle n \mid  [ \hat{S}_+(i) , \hat{S}_-(j) ] \mid  n \rangle  $.

\appendix



\end{document}